\documentclass[conference, twocolumn]{IEEEtran}

\usepackage[utf8]{inputenc}
\usepackage{microtype}

\usepackage{amssymb}
\usepackage[abs]{overpic}
\usepackage{graphicx}
\usepackage{subcaption}
\usepackage{siunitx}
\usepackage[hyphens]{url}
\usepackage{cite}
\usepackage{paralist}
\usepackage{enumerate}
\usepackage{enumitem}
\usepackage{algorithm}
\usepackage{xspace}
\usepackage[noend]{algpseudocode}

\usepackage{booktabs}
\usepackage[table,xcdraw,dvipsnames]{xcolor}

\usepackage[flushleft]{threeparttable}

\usepackage{flushend}
\usepackage{wasysym}%
\newcommand{\bl}{\blacksquare}
\newcommand{\sq}{\square}

\usepackage{balance}
\usepackage{dcolumn}
\usepackage{longtable,tabu}
\usepackage{threeparttablex}
\usepackage{adjustbox}
\usepackage{tabularx}
\usepackage{tikz}
\usepackage{pgfplots}
\pgfplotsset{width=6.5cm,compat=1.9}
\usepgfplotslibrary{statistics}

\usepackage[underline=true]{pgf-umlsd}
\usetikzlibrary{calc}
\usetikzlibrary{arrows.meta}

\usepackage{arydshln}
\usepackage{multirow}

\usepackage{pifont}
\usepackage{enumitem}
\newcommand{\subscript}[2]{$#1 _ #2$}

\usepackage[mathlines,switch]{lineno}

\begin{document}
\title{Fair and Transparent Blockchain based Tendering Framework - A Step Towards Open Governance}
\author{}
\author{\IEEEauthorblockN{Freya Sheer Hardwick, Raja Naeem Akram, and Konstantinos Markantonakis}
\IEEEauthorblockA{ISG-SCC, Royal Holloway, University of London, Egham, United Kingdom\\
Email: \{Freya.SheerHardwick.2016\}@live.rhul.ac.uk, \{r.n.akram, k.markantonakis\}@rhul.ac.uk}}

\maketitle

\begin{abstract}
At a time when society is in constant transition to keep up with technological advancement, we are seeing traditional paradigms being increasingly challenged. 
The fundamentals of governance are one such paradigm. As society’s values have shifted, so have expectations of government shifted from the traditional model to something commonly referred to as ‘open governance’. 
Though a disputed term, we take open governance to mean a concept, which encourages and facilitates openness, accountability, and responsiveness to citizens. 
For the success of open governance initiatives, there are some technologies, such as the internet, that are crucial. 
These technologies enable access to both the data and to engagement activities between citizens and government. 
There are also other technologies, like blockchain and smart contacts, which could be utilised to assist open governance. 
A sound starting point would be moving from a system where information is tediously released by a government, on an ‘as they please’ basis, to an infrastructure where critical actions are captured with strong integrity, non-repudiation and evidential guarantees.
With an added dimension that facilitates these actions record be accessible to public scrutiny in near real-time. 
One candidate technology for capturing such actions is blockchain. Initially, blockchains were mainly used to facilitate cryptocurrencies as a record of transactions. 
The notable example being bitcoin. 
However, in recent years, blockchains utility is being recognised through smart contracts - potentially a vital building block to realising open and transparent government activities. 
In this paper, we employ the concept of smart contracts to government tendering activities. 
The proposed scheme is based on smart contracts, enabling a fair, transparent and independently verifiable (auditable) government tendering scheme. 
The scheme is then implemented on the Ethereum platform to evaluate the performance and financial cost implications, along with an evaluation of the potential security and auditability challenges. 
\end{abstract}

\IEEEpeerreviewmaketitle

\section{Introduction}
\label{sec:Introduction}
The phrase ‘fair and open’ has different meanings depending on the context and situation. 
It is more so right for government where the phrase is regularly applied. 
Fair and open means that the government do not act with bias and on request from a citizen it shares the necessary information (Right to Information \cite{davis1967information}). 
This does not mean that openness translates into trust in governance \cite{worthy2010more}. 
The issue with this model is two-fold; \begin{inparaenum}[\itshape a)\upshape]
\item The access to information is cumbersome and requires a long process to get hold of information. Even for the information that has no national security implications,
\item even though auditing the government activities might be possible but still reviewing the documents (got through Right to Information) requires time, money and expertise -- something not easy to acquire/invest by the general public.  
\end{inparaenum}

On the other hand, we have e-government initiatives \cite{heeks2007analyzing} that enables the use of technology in government activities \cite{MCDERMOTT2010401}, fostering transparency, participation, and collaboration between the citizens and government. 
An amalgamation of these two: open governance and e-government, supported by innovative technologies like blockchain, has the potential to introduce fairness, openness, and accountability. 
Thus enabling an independent and automatic auditing process to provide accountability and allowing citizens to track the activities of their government without unnecessary hassle. 

\textit{The paper proposes that the blockchain and smart contracts can enable an open governance framework that can facilitate citizens oversight on government functions that is easy to carry out with no associated financial costs.}

Therefore as a proof-of-concept, we explore the government tendering process, a set of activities that have three distinct phases: \begin{inparaenum}[\itshape a)\upshape]
\item government tender opening (publishing),
\item bidding period, and 
\item tender closing and selection of the best bid.
\end{inparaenum}

Morphing these activities as part of our proposal in such a manner that it enables;

\begin{enumerate}
\item The tendering organisation (like a government) can open a tender, and once the tender is open, they cannot change it. 
Preventing the organisation from changing the tender to favour a bidding organisation (e.g. government departments or commercial organisations). 
Furthermore, each tender includes evaluation criteria for selecting the best possible bid. 
\item Authorised bidding organisations can place a bid with an assurance that their bid is confidential (until the tendering closing date/time) and will not be modified (integrity protection). 
Also, there is some assurance that third parties would not be able to place bids on behalf of other authorised bidding organisations. 
Furthermore, during the bidding process, individual bidding organisation cannot know what bid the other organisations have placed. 
Also, as a stringent privacy requirement, bidding organisations should not find out whether a particular organisation has placed a bid or not. 
\item The tendering organisation can only open the bids after the tender is closed. 
The selection of the best bid would be published. 
Organisations losing the bid can compare the winning bid with theirs to evaluate the decisions -- using the evaluation criteria. 
Furthermore, if required, all bids for the tender can be made public and so citizens\footnote{In the context of the paper, citizens are defined as the general public that politically belongs to or shareholder in a government.} or other interested parties can also evaluate the whole bidding process. 
\item The technology provides non-repudiation, collision avoidance, confidentiality (time-dependent), privacy and integrity -- along with independent auditability feature and evidential guarantees.
\end{enumerate}

\subsection{Paper's Contributions}
\label{sec:Papers_Contributions}
The contributions of the paper can be summarised as; \begin{inparaenum}[\itshape C-i)\upshape]
\item A generic architecture to deploy open tendering scheme using blockchain, 
\item Three proposed deployment variants of the open tendering scheme, and 
\item Implementation details, performance evaluations, and security analysis of the proposed deployment variants of the open tendering scheme.
\end{inparaenum}

\begin{figure}[htpb]
 \centering
 \includegraphics[width=\columnwidth]{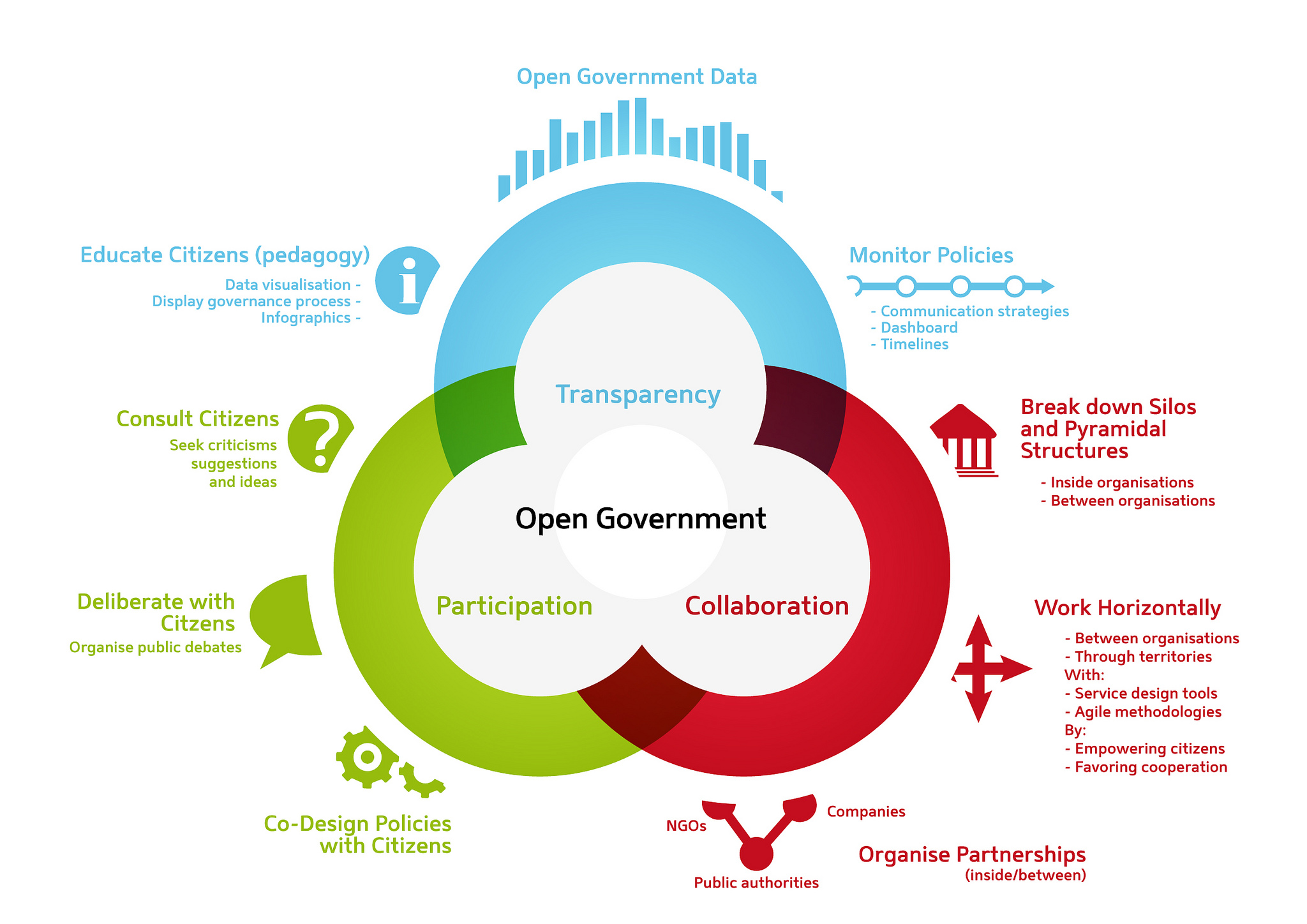}
 \caption{Open Governance Diagram by Armel Le Coz and Cyril Lage (under creative commons attribution terms)}
 \label{fig:OpenGovernanceDiagram} 
\end{figure}

\section{Blockchain Technology and Transparent Governance}
\label{sec:BlockChainTechandTransGover}
In this section, we first briefly discuss the open governance initiatives followed by a succinct description of the government tendering process.  
Finally, we conclude the section with a short explanation of blockchain technology and smart contracts. 

\subsection{Open Governance}
\label{sec:OpenandTransGover}
According to the Open Government Partnership\footnote{A global initiative with participation from 71 countries either through national or local action. 
Website \url{www.opengovpartnership.org}}, open government is about making national and/or local authorities promote transparency, participation, and collaboration \cite{MCDERMOTT2010401} - as represented in Figure \ref{fig:OpenGovernanceDiagram}. 
Furthermore, to achieve these goals, they are encouraged to harness new technologies to strengthen governance \cite{lee2012open}. 
In recent years, the focus of open government has shifted from political activities to more emphasis on information/data and citizen-centric services \cite{yu2011new}. 
The term used is Open Data defined as data related to government operations, but not sensitive to national security and accessible to anyone. 
Example of such an initiative is UK governments open data portal \url{https://data.gov.uk} that provides a single point to access government's public data. 
There are some interesting and useful mobile Apps based on the UK government's public data. 
For example, CheckMyStreet\footnote{\url{https://www.checkmystreet.co.uk}} (information about local properties, local amenities, average monthly rents and local crime statistics), Regisearch\footnote{\url{https://regisearch.co.uk}} (vehicle information, prices and MOT history) and Traffic Injuries Map\footnote{\url{http://www.road-injuries.info/map.html}}.

One significant distinction we would like to point out is that not all open data initiatives even come close to open government initiatives. 
A government can make its data public as it deems appropriate and open data, in many cases, does not equate to being transparent and nor does it necessarily encourage citizen participation. 
Without the notion of accountability, open data (and open governance) cannot have the desired impact.

\subsection{Government Tendering Framework}
\label{sec:GovernmentTenderingFramework}
\begin{figure}[ht]
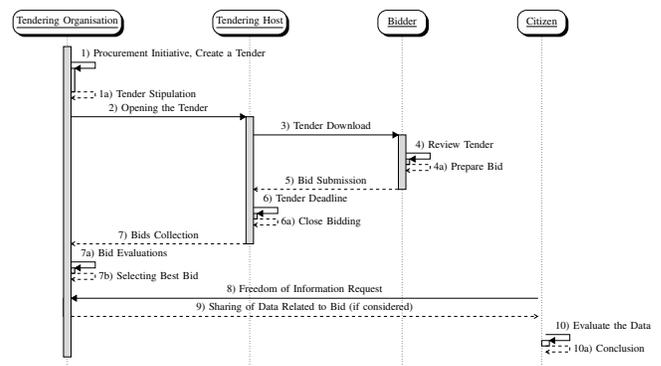

 \centering
 \resizebox{\columnwidth}{!}{%
  \begin{sequencediagram}
   \tikzstyle{inststyle}+=[rounded corners=3mm]
   \newthread{TeOr}{Tendering Organisation}
   \newinst[3]{TeHo}{Tendering Host}
   \newinst[3]{Bidder}{Bidder}
   \newinst[3]{Citizen}{Citizen}
            
            \begin{callself}{TeOr}{1) Procurement Initiative, Create a Tender}{1a) Tender Stipulation}
             \postlevel
            \end{callself}
            
            \begin{call}{TeOr}{2) Opening the Tender }{TeHo}{7) Bids Collection}
             \begin{call}{TeHo}{3) Tender Download}{Bidder}{5) Bid Submission}
                 \begin{callself}{Bidder}{4) Review Tender }{4a) Prepare Bid}
                    \end{callself}
             \end{call}    
                \begin{callself}{TeHo}{6) Tender Deadline}{6a) Close Bidding} 
                \end{callself}
            \end{call}
            
            \begin{callself}{TeOr}{7a) Bid Evaluations}{7b) Selecting Best Bid}
            \end{callself}
            
            \begin{call}{Citizen}{8) Freedom of Information Request}{TeOr}{9) Sharing of Data Related to Bid (if considered)}
            \end{call}
            \begin{callself}{Citizen}{10) Evaluate the Data}{10a) Conclusion}
            \end{callself}
  \end{sequencediagram}
 }
 \caption{Generic Tendering Framework}
 \label{fig:GTFramework}
\end{figure}

The actual process for government procurement of services and products depends on individual governments or geographical-zones. 
In this section, we describe a generic framework that would explain the unique steps taken during a procurement process. 
Figure \ref{fig:GTFramework}, shows the procurement process that is explained as below.

\begin{enumerate}
\item Based on the services and product requirements, a tendering organisation (i.e., respective government department) would initiate the process of putting out a request for tender. 
 \begin{enumerate}[label=a)]
 \item A tender specification would include the terms and conditions of the tender, information necessary for an acceptable bid, and bid evaluation criteria. 
 \end{enumerate}
\item The tender with its full specification would be published through the tendering host. The host can be a separate department in a government or part of the tendering organisation. They publicise the tender specification over the internet, in newspapers, and in relevant industry news magazines/portals. 
\item An interested bidding organisation would download/access the tender specification
\item The respective bidding organisation would review the requirements.
 \begin{enumerate}[label=a)]
 \item Based on the tender specifications, the bidding organisation would prepare a bid. 
 \end{enumerate}
\item The prepared bid is submitted to the tendering host
\item Submission of the bids would be open for a limited period -- depending upon the tender specification. 
 \begin{enumerate}[label=a)]
 \item When the deadline has passed, the tender host will shut down the bid submission portal. All bids received after this point would be rejected. 
 \end{enumerate}
\item Tendering organisations will evaluate all of the submitted bids as per the evaluation criteria stipulated in the tender specification. 
 \begin{enumerate}[label=a)]
 \item Based on the evaluation, the best bid would be selected and notified by the tendering organisation.
 \end{enumerate}
\item From step 1 to 7, citizens are not involved and have no visibility. However, after the tender is concluded, they can request for the data associated with the respective tendering process. 
\item If a government deems it appropriate, they can fulfil the request and provide the requester with the data. Under open data initiatives, some governments would even make its data available voluntarily. 
\item Citizens have to be resourceful\footnote{The notion of resourceful means that citizens have both financial resources and required skills to interpret the data and conduct an audit to find out any discrepancies} in the evaluation of the data/process to make use of this data. 
 \begin{enumerate}[label=a)]
 \item Based on this evaluation they can judge whether the process was fair or not. A bidding organisation can do the same, just as any citizen. One point to note is that the process of getting data and then evaluating it is so laborious that it is seldom taken up by a single individual. Moreover, most of the time, such activities never go through the scrutiny of the general public as an examination has a substantial cost in both time and money. 
 \end{enumerate}

\end{enumerate}

The proposition of this paper is that we can design a secure, fair, and reliable bidding architecture that can transparently manage the whole bidding process and allow a citizen to evaluate the process with a single click through an auditing application/portal. 
Through this, we can achieve higher participation of citizens in government activities and also increase government transparency and accountability. 

\subsection{BlockChain Technology}
\label{sec:BlockchainTechnology}
A blockchain is a form of a distributed database \cite{454b7bf19bef4c}. 
A distributed database is a collection of interrelated databases stored in multiple locations. 
In most traditional senses of the term, a distributed database is divided into portions that are then maintained in these separate locations \cite{ozsu2011principles}. 
In the case of the blockchain, every participating node has a copy of the ‘database' in its entirety. 
Meaning that the participating nodes do not need to trust each other to trust the data stored in the ledger \cite{beck2016blockchain,yin2017m2m}. 
Each node can independently verify the data they are given and then decide to store that data in their copy of the database. 
This results in a database that grows by consensus.

Transactions that are intended to be added to the blockchain are propagated through the network by the participating nodes \cite{Jacobs:2017}. 
To mine a block, blockchains employ a consensus protocol \cite{cachin2016architecture} that must be satisfied by the block (one such example is the proof-of-stake protocol) Once a block is mined, this new block is broadcast to the other nodes, which will then append it to their active chain. 
These blocks will have a cryptographic hash in the header that relates to the previous block in the chain. 
Sometimes, two nodes will mine two different blocks at the same time, and both broadcast these blocks creates a fork in the chain. 
Nodes resolve this by always selecting the chain with the most work performed on it as their active chain. 
Resulting in the chain being supported by the majority of the nodes always becoming the active node. 
Thus consensus of more than 50\% of the nodes prevails. 
Ensuring the integrity of the chain, together with the consensus protocol and the cryptographic hashes. 
The general concept can only be spoken about vaguely because how this plays out in practice depends entirely on the implementation of the blockchain and the consensus protocol employed.

\subsection{Smart Contracts}
\label{sec:SmartContracts}
A smart contract is a piece of self-executing code that can be stored, and executed, on the blockchain \cite{kosba2016hawk}.

A smart contract is deterministic, verifiable, and doesn't rely on any trusted third party \cite{bhargavan2016formal}. 
Entities can enter into an agreement with all of the terms transparent to them. 
The same integrity checks that keep the transactions on the blockchain from being edited are also in effect here. 
This means that when entities enter into the agreement, they can be sure that no party will edit the terms of that agreement at a later date. 

Smart contracts also have state and memory storage and so can hold assets in their own right \cite{luu2016making}. 
Implying that they can be used to hold funds in escrow in instances of asset transfer between parties. 
The applicability of this goes far beyond the crypto-currencies that are currently popularising the blockchain. 

The limitations of smart contracts are entirely in the expressiveness of the language supported by the blockchain. 
With a Turing complete language, as is employed by Ethereum, smart contracts can be used to execute a number of functions. 
Therefore, smart contracts provide a trustless environment for asset exchange.

\section{Open and Transparent Tendering Framework}
\label{sec:OpenandTransTendFrame}
In this section, we will briefly describe the open and transparent tendering framework based on blockchain and smart contract technologies. 

\subsection{Overall Architecture}
\label{sec:OverallArchitecture}
\begin{figure}[ht]
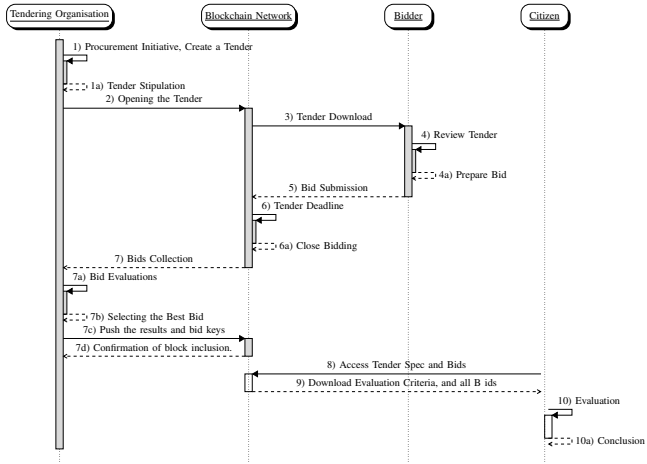

 \centering
 \resizebox{\columnwidth}{!}{%
  \begin{sequencediagram}
   \tikzstyle{inststyle}+=[rounded corners=3mm]
   \newthread{TeOr}{Tendering Organisation}
   \newinst[3]{TeHo}{Blockchain Network}
   \newinst[3]{Bidder}{Bidder}
   \newinst[3]{Citizen}{Citizen}
            
            \begin{callself}{TeOr}{1) Procurement Initiative, Create a Tender}{1a) Tender Stipulation}
             \postlevel
            \end{callself}
             
            \begin{call}{TeOr}{2) Opening the Tender }{TeHo}{7) Bids Collection}
             
             \begin{call}{TeHo}{3) Tender Download}{Bidder}{5) Bid Submission}
             
                 \begin{callself}{Bidder}{4) Review Tender }{4a) Prepare Bid}\postlevel
                    \end{callself}
                
             \end{call}    
                
                \begin{callself}{TeHo}{6) Tender Deadline}{6a) Close Bidding}\postlevel
                
                \end{callself}
                
            \end{call}
            
            \begin{callself}{TeOr}{7a) Bid Evaluations}{7b) Selecting the Best Bid}\postlevel
            \end{callself}
            
           \begin{call}{TeOr}{7c) Push the results and bid keys}{TeHo}{7d) Confirmation of block inclusion.}
           \end{call}
            
            \begin{call}{Citizen}{8) Access Tender Spec and Bids}{TeHo}{9) Download Evaluation Criteria, and all B ids}
             
            \end{call}
            \begin{callself}{Citizen}{10) Evaluation}{10a) Conclusion}\postlevel
            \end{callself}

  \end{sequencediagram}
 }
 \caption{Smart Contract Based Tendering Architecture}
 \label{fig:SCFramework}
\end{figure}

Figure \ref{fig:SCFramework} depicts the inclusion of the blockchain and smart contract technologies to the tendering framework - discussed in Section \ref{sec:GovernmentTenderingFramework}. We are only going to discuss the steps that are different from the generic tendering framework (Figure \ref{fig:GTFramework}).

\begin{enumerate}
\setcounter{enumi}{0}
\item A tendering organisation will create a tender as a smart contract and place it on the blockchain. The smart contract will include the certified public key\footnote{For security reasons, state-of-the-art public key cryptosystem \cite{menezes2012elliptic} should be used with recommended configuration and key sizes as suggested by competent authorities like NIST. } ($P_{TO}$) of the tendering organisation along with bid evaluation code.
\item A prospective bidder can download the tender from the blockchain. 
\item The respective bidder reviews the tender and 
 \begin{enumerate}[label=\alph*)]
 \item Consider the tendering specification and make a bid proposal.
 \item Generates a bid in response to the tender (smart contract). The actual bid is encrypted by the bidder's generated symmetric key\footnote{Similar to the public key recommendation, for symmetric key standard schemes like Advanced Encryption Standard should be used with adequate key size (e.g. AES 256) \cite{daemen2013design}.} (bid key: $SK_{Bidder}$). The symmetric key is then encrypted by the public key of tendering organisation: $P_{TO}(SK_{Bidder})$. Half of the $P_{TO}(SK_{Bidder})$ is included as part of the submission and the second half would be communicated to the tendering organisation at the tender submission deadline. 
 \end{enumerate}
\item The bidder will push the bid as a smart contract to the blockchain. The bid is signed by the bidder's certified signature key. This key is certified by the tendering organisation when the bidder register as an authorised bidding company - a process out of the actual tender opening and allocation process. 
\item When the deadline for bid submission expires, the smart contract on the blockchain stops accepting new bids. 
\item The tendering organisation can download the submitted bids, and they can decrypt the bids if they have full $P_{TO}(SK_{Bidder})$.
 \begin{enumerate}[label=\alph*)]
 \item At the tender closing date, tendering organisation will run the evaluation code and select the best bid. 
 \item The result of the evaluation is pushed to the blockchain. At this stage, the tendering organisation can make $P_{TO}(SK_{Bidder})$ of all bidders public on the blockchain. 
 \end{enumerate}
 \item The tender organisation will push the results of the bid evaluations along with bidder's keys to the blockchain. This information is crucial for independent auditing of the tendering process.
\item Citizens can access the tender details from the blockchain (where this data will reside in perpetuity) along with the bid evaluation code
\item Citizens can download the tender contract that contains the code for bid evaluation criteria. 
 \begin{enumerate}[label=\alph*)]
 \item Citizens just have to run the evaluation code that will read the bids from the block and evaluation them. 
 \item The results of the evaluation will show whether the bidding process was fair (auditing tender allocation to the stated best bidder). 
 \end{enumerate}

\end{enumerate}
\begin{table*}[t]
\begin{minipage}[b]{0.32\textwidth}
\centering
 \begin{tabular}{||c c c||} 
 \hline
 Contract & Time (secs) & Cost (gas) \\ [1ex] 
 \hline\hline
 1 & 198.57 & 892160 \\ 
 \hline
 2 & 88.86 & 892160  \\
 \hline
 3 & 124.34 & 892160  \\
 \hline
 4 & 96.52 & 892160 \\
 \hline
 5 & 103.65 & 892160 \\  
 \hline
 6 & 174.69 & 892160 \\ 
 \hline
 7 & 166.33 & 892160  \\
 \hline
 8 & 137.11 & 892160  \\
 \hline
 9 & 133.56 & 892160 \\
 \hline
 10 & 253.98 & 892160 \\ 
 \hline\hline
 Average & 147.761 & 892160 \\ 
 \hline
\end{tabular}
   \caption{Full Track Scheme: Contract Deployment}
    \label{table:FTContractDeployment}
\end{minipage}\hspace{2mm}
\begin{minipage}[b]{0.33\textwidth}
\centering
\begin{tikzpicture}
\begin{axis}[
    ylabel={Gas Consumption in 10$^5$.},
    xmin=0, xmax=11,
    ymin=250000, ymax=500000,
    xtick={1,2,3,4,5,6,7,8,9,10},
    ytick={250000,300000,350000,400000,450000,500000,550000},
    legend pos=north west,
    ymajorgrids=true,
    grid style=dashed,
]
\addplot[
    color=blue,
    mark=square,
    ]
    coordinates {
    (1,299501)(2,320282)(3,341063)(4,361844)(5,382626)(6,403407)(7,424191)(8,444975)(9,465759)(10,486607)
    };
    \legend{Bid's Gas Consumption}
\end{axis}
\end{tikzpicture}

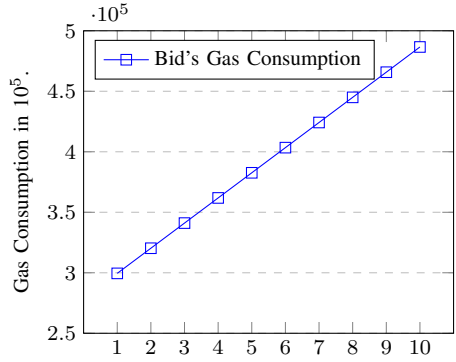
\captionof{figure}{Individual Bid's GAS usage (x 10$^5$) for Full Track Scheme}
\label{fig:FTContractGasUsage}
\end{minipage}\hspace{2mm}
\begin{minipage}[b]{0.33\textwidth}
\centering
\begin{tikzpicture}
  \begin{axis}
    [ylabel = {Time (Seconds)},
    boxplot/draw direction=y,
    xtick={1,2,3, 4, 5, 6, 7, 8, 9, 10},
    xticklabels={T1, T2, T3, T4, T5, T6, T7, T8, T9, T10},
    x tick label style={font=\footnotesize, text width=2.5cm, align=center}
    ]
    \addplot+[
    boxplot prepared={
      lower whisker=15,
      lower quartile=23.5,
      median=29,
      upper quartile=32.75,
      upper whisker=57,
      average=31.9
    }, color = red
    ]coordinates{};
    \addplot+[
    boxplot prepared={
      lower whisker=18,
      lower quartile=24.5,
      median=27.5,
      upper quartile=30.25,
      upper whisker=60,
      average=29.7
    }, color = blue
    ] coordinates{};
    \addplot+[
    boxplot prepared={
      lower whisker=18,
      lower quartile=23.75,
      median=29.5,
      upper quartile=44,
      upper whisker=56,
      average=43.2
    }, color = Aquamarine
    ] coordinates{};
    \addplot+[
    boxplot prepared={
      lower whisker=12,
      lower quartile=14.5,
      median=22.5,
      upper quartile=36.25,
      upper whisker=56,
      average=27.2
    }, color = Orange
    ]coordinates{};
    \addplot+[
    boxplot prepared={
      lower whisker=19,
      lower quartile=23.75,
      median=43.5,
      upper quartile=56,
      upper whisker=88,
      average=43.4
    }, color = OrangeRed
    ] coordinates{};
    \addplot+[
    boxplot prepared={
      lower whisker=9,
      lower quartile=21.75,
      median=31.5,
      upper quartile=34.75,
      upper whisker=37,
      average=28,
      every box/.style={thin,solid,draw=Periwinkle,fill=white},
      every whisker/.style={Periwinkle, thin, solid},
      every median/.style={solid,Periwinkle,thin}
    }, color = Periwinkle
    ] coordinates{};    
    \addplot+[
    boxplot prepared={
      lower whisker=13,
      lower quartile=31.75,
      median=42,
      upper quartile=56.75,
      upper whisker=120,
      average=49,
      every box/.style={thin,solid,draw=Purple,fill=white},
      every whisker/.style={Purple, thin, solid},
      every median/.style={solid,Purple,thin}
    }, color = Purple
    ]coordinates{};
    \addplot+[
    boxplot prepared={
      lower whisker=17,
      lower quartile=23.25,
      median=36.5,
      upper quartile=43.75,
      upper whisker=60,
      average=35.5,
      every box/.style={thin,solid,draw=CarnationPink,fill=white},
      every whisker/.style={CarnationPink, thin, solid},
      every median/.style={solid,CarnationPink,thin}
    }, color = CarnationPink
    ] coordinates{};
    \addplot+[
    boxplot prepared={
      lower whisker=17,
      lower quartile=22.5,
      median=24.5,
      upper quartile=40,
      upper whisker=82,
      average=33.9,
      every box/.style={thin,solid,draw=Maroon,fill=white},
      every whisker/.style={Maroon, thin, solid},
      every median/.style={solid,Maroon,thin}
    }, color = Maroon
    ] coordinates{};
       \addplot+[
    boxplot prepared={
      lower whisker=10,
      lower quartile=31.25,
      median=46.5,
      upper quartile=68,
      upper whisker=78,
      average=47.2,
      every box/.style={thin,solid,draw=JungleGreen,fill=white},
      every whisker/.style={JungleGreen, thin, solid},
      every median/.style={solid,JungleGreen,thin}
    }, color = JungleGreen
    ] coordinates{};
    \end{axis}
\end{tikzpicture}


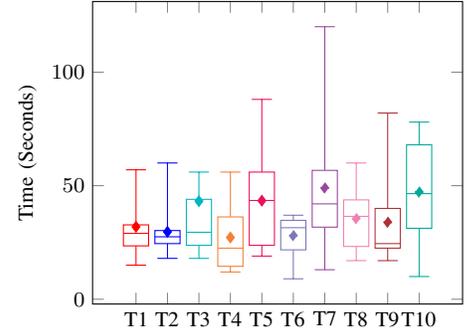
\captionof{figure}{Contract timing for Full Track Scheme Over Ten Trial Tenders}
\label{fig:FTimage}
\end{minipage}
\end{table*}


\subsection{Security and Operational Requirements}
\label{sec:SecandOperReq}
In this section, we highlight the security and operation requirements any implementations of the open and transparent tendering framework has to satisfy.
\begin{enumerate}[label=\subscript{R}{{\arabic*}})]
\item The tendering Organisation cannot change the tender once it is placed on the blockchain. If due to some unforeseeable reasons they have to change it, then they have to create a new tender (smart contract) on the blockchain. 
\item The tendering organisation cannot read the bid until the deadline is expired. 
\item Bidders cannot change the bids of other organisation.
\item Bidders cannot see who else has placed a bid.
\item Bidders cannot mount a Denial-of-Service (DOS) attack on their competitors to stop competitors placing a bid on the blockchain. 
\item Blockchain network or block miners cannot affect the tendering process.
\end{enumerate}

\section{Implementation and Operational Evaluation}
\label{sec:ImplementationandOperationEvaluation}
In this section, we will detail the implementation of the proposed architecture - with it three variants: a) Full Track Scheme, b) Protected State Scheme and c) Stateless Scheme. We also list the performance costs and blockchain network usage costs (GAS). 

\subsection{Implementation Details}
\label{sec:ImplementationDetails}
In the implementation of the tender framework, we have used the Ethereum blockchain API. We made this choice because Ethereum is an open-source platform that is publicly available and a well-known choice for developing distributed applications.

With Ethereum, each transaction will have a gas usage and gas cost. The gas usage is determined by how computationally expensive the transaction is. The purpose of working out the gas used in each transaction comes down to the fact that every node in the blockchain verifies the transaction. If transactions were allowed to be arbitrarily complex, verification on the network would be slow and result in a processing bottleneck. Citing gas needed for each transaction allows miners to determine whether or not it is worth including the transaction in the block that they are mining and they will use the cost of gas, as set on each transaction by the node that pushed it, to determine this. Trying to push a transaction that is too complex or with the gas cost set too low will cause a transaction to be ignored by the miners when they pick transactions to include in their block.

To mine one of these blocks, a node must satisfy the proof-of-work constraint (Ethereum's choice for the consensus protocol)  \cite{gervais2016security}. This constraint necessitates a certain degree of work to be undergone by the node that wishes to push a block to the wider chain. This stops a node from pushing an arbitrary number of blocks to the chain: it would be computationally unfeasible.

For a malicious node to corrupt a block halfway down the chain and present it as the valid active chain, it would also need to recompute every following block with their new cryptographic hash faster than every other participating node is working on their active chains.

We used Truffle JS \cite{wimmer2012truffle} as our framework to create our contracts and wrote them in the Solidity language. For the unit tests during development, we used a private blockchain running on the TestRPC Ethereum client, and our operational performance tests were done using the Ropsten testing network. We chose this network because online documentation would suggest that this environment most closely resembled the Ethereum production environment at the time of testing. To connect to Ropsten, we used the infura.io node. The blockchain explorer we used to interact with the contracts is MyEtherWallet using an account made via the MetaMask chrome extension.\\

\begin{algorithm}
\caption{Initiating A Tender}\label{alg:initiateTender}
\addtolength\linewidth{-4ex}
\small
\begin{algorithmic}[1]
\Procedure{ReqForTender}{$\_length,\_pubk,\_limit$}
\State $biddingEnd\gets TimeNow() +  \_length$
\State $limit\gets \_limit$
\State $pubk\gets \_pubk$
\EndProcedure
\end{algorithmic}
\end{algorithm}

The Request For Tender (Algorithm \ref{alg:initiateTender}) is initiated by a contract placed on the blockchain. This contract is created with a length, given in milliseconds, that determines how long the contractors (also referred to as Tendering Organisation: TO) have to place their bids. This time is calculated using the Unix Epoch time at the time of creation. An upper limit is also given, which will be used to control the number of tenders a contractor can place for this auction. The entity that created the auction is also required to pass in a public key that is specific to this request for tender contract. These three attributes are held in the contract's state.

\begin{table*}[t]
\begin{minipage}[b]{0.32\textwidth}
\centering
 \begin{tabular}{||c c c||} 
 \hline
 Contract & Time (secs) & Cost (gas) \\ [1ex] 
 \hline\hline
 1 & 118.55 & 874791 \\ 
 \hline
 2 & 114.3 & 874791  \\
 \hline
 3 & 118.52 & 874791  \\
 \hline
 4 & 100.18 & 874791 \\
 \hline
 5 & 87.77 & 874791 \\  
 \hline
 6 & 89.6 & 874791 \\ 
 \hline
 7 & 70.39 & 874791  \\
 \hline
 8 & 122.03 & 874791  \\
 \hline
 9 & 61.77 & 874791 \\
 \hline
 10 & 99.82 & 874791 \\ 
 \hline\hline
 Average & 98.293 & 874791 \\ 
 \hline
\end{tabular}
\caption{Protected State Schema: Contract Deployment}
\label{table:ProStateSchContractDeployment}
\end{minipage}\hspace{2mm}
\begin{minipage}[b]{0.33\textwidth}
\centering
\begin{tikzpicture}
\begin{axis}[
    ylabel={Gas Consumption in 10$^5$.},
    xmin=0, xmax=11,
    ymin=300000, ymax=550000,
    xtick={1,2,3,4,5,6,7,8,9,10},
    ytick={300000,350000,400000,450000,500000,550000},
    legend pos=north west,
    ymajorgrids=true,
    grid style=dashed,
]
\addplot[
    color=blue,
    mark=square,
    ]
    coordinates {
    (1,332788)(2,353569)(3,374350)(4,395131)(5,415913)(6,436694)(7,457478)(8,478262)(9,499046)(10,519830)
    };
    \legend{Bid's Gas Consumption}
\end{axis}
\end{tikzpicture}

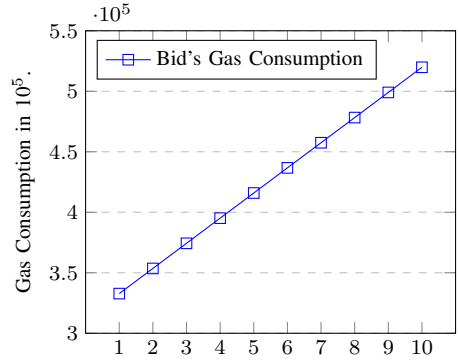
\captionof{figure}{Individual Bid's GAS usage (x 10$^5$) for Protected Scheme}
\label{fig:PSContractGasUsage}
\end{minipage}\hspace{2mm}
\begin{minipage}[b]{0.33\textwidth}
\centering
\begin{tikzpicture}
  \begin{axis}
    [ylabel = {Time (Seconds)},
    boxplot/draw direction=y,
    xtick={1,2,3, 4, 5, 6, 7, 8, 9, 10},
    xticklabels={T1, T2, T3, T4, T5, T6, T7, T8, T9, T10},
    x tick label style={font=\footnotesize, text width=2.5cm, align=center}
    ]
    \addplot+[mark = *, mark options = {black}, 
    boxplot prepared={
      lower whisker=20,
      lower quartile=27.5,
      median=31.5,
      upper quartile=35.5,
      upper whisker=72,
      average=34.6
    }, color = red
    ]coordinates{};
    \addplot+[mark = *,mark options = {blue},
    boxplot prepared={
      lower whisker=8,
      lower quartile=19.75,
      median=25.5,
      upper quartile=39.25,
      upper whisker=47,
      average=28.3
    }, color = blue
    ] coordinates{};
    \addplot+[mark = *,mark options = {cyan}, 
    boxplot prepared={
      lower whisker=26,
      lower quartile=35.25,
      median=46.5,
      upper quartile=48.75,
      upper whisker=67,
      average=43.8
    }, color = Aquamarine
    ] coordinates{};
    \addplot+[mark = *, mark options = {Aquamarine}, 
    boxplot prepared={
      lower whisker=19,
      lower quartile=21.5,
      median=31.5,
      upper quartile=42.75,
      upper whisker=64,
      average=34.4
    }, color = Orange
    ]coordinates{};
    \addplot+[mark = *,mark options = {Orange}, 
    boxplot prepared={
      lower whisker=12,
      lower quartile=25,
      median=29,
      upper quartile=34.75,
      upper whisker=65,
      average=31.1
    }, color = OrangeRed
    ] coordinates{};
    \addplot+[mark = *,mark options = {OrangeRed}, 
    boxplot prepared={
      lower whisker=13,
      lower quartile=22.25,
      median=31.5,
      upper quartile=36,
      upper whisker=48,
      average=30.1,
      every box/.style={thin,solid,draw=Periwinkle,fill=white},
      every whisker/.style={Periwinkle, thin, solid},
      every median/.style={solid,Periwinkle,thin}
    }, color = Periwinkle
    ] coordinates{};    
    \addplot+[mark = *, mark options = {red}, 
    boxplot prepared={
      lower whisker=12,
      lower quartile=27.75,
      median=34,
      upper quartile=42.5,
      upper whisker=61,
      average=35.9,
      every box/.style={thin,solid,draw=Purple,fill=white},
      every whisker/.style={Purple, thin, solid},
      every median/.style={solid,Purple,thin}
    }, color = Purple
    ]coordinates{};
    \addplot+[mark = *,mark options = {blue}, 
    boxplot prepared={
      lower whisker=12,
      lower quartile=22.25,
      median=29,
      upper quartile=51.25,
      upper whisker=56,
      average=34.2,
      every box/.style={thin,solid,draw=CarnationPink,fill=white},
      every whisker/.style={CarnationPink, thin, solid},
      every median/.style={solid,CarnationPink,thin}
    }, color = CarnationPink
    ] coordinates{};
    \addplot+[mark = *,mark options = {green}, 
    boxplot prepared={
      lower whisker=15,
      lower quartile=18.5,
      median=28.5,
      upper quartile=36.5,
      upper whisker=56,
      average=30,
      every box/.style={thin,solid,draw=Maroon,fill=white},
      every whisker/.style={Maroon, thin, solid},
      every median/.style={solid,Maroon,thin}
    }, color = Maroon
    ] coordinates{};
    \addplot+[mark = *,mark options = {green}, 
    boxplot prepared={
      lower whisker=18,
      lower quartile=25,
      median=31,
      upper quartile=44,
      upper whisker=78,
      average=37.3,
      every box/.style={thin,solid,draw=JungleGreen,fill=white},
      every whisker/.style={JungleGreen, thin, solid},
      every median/.style={solid,JungleGreen,thin}
    }, color = JungleGreen
    ] coordinates{};
    \end{axis}
\end{tikzpicture}

\captionof{figure}{Bid Timing for Protected Scheme Over Ten Trial Tenders}
\label{fig:PSimage}
\end{minipage}
\end{table*}


\begin{algorithm}
\caption{Placing A Bid (Full Track Scheme)}\label{alg:placeBidOS}
\addtolength\linewidth{-4ex}
\small
\algnewcommand{\And}{\textbf{and}\xspace}
  \begin{algorithmic}[1]
   \Procedure{PlaceBid}{$id,data,msgHashed,v,r,s,$}
    \State $bidValidity\gets ValidBid(id,msgHashed,v,r,s)$
    \If{$bidValidity$}
     \State $bidCount[id] += 1$
    \EndIf
    \State \emph{bid}$\gets$ new \emph{Bid}     
             (\emph{id},\emph{data},\emph{bidValidity},\emph{bidsPlaced},\emph{biddingEnd})
    \State $bidsPlaced.add(bid)$
    \State \textbf{return} $bid$
   \EndProcedure
   \Procedure{ValidBid}{$id,msgHashed,v,r,s,$}
    \State $validHash\gets verify(msgHash,v,r,s)$ 
    \State $validTime\gets timeNow() < biddingEnd$
    \State $allowedBid\gets bidCount[id] < limit$
    \State \textbf{return} $validHash$ \And $validTime$ \And $allowedBid$
   \EndProcedure
   \Procedure{Bid}{\emph{\_id},\emph{\_data},\emph{\_validity},\emph{\_bidsPlaced},\emph{\_biddingEnd}}
    \State $id\gets \_id$ 
    \State $data\gets \_data$
    \State $validity\gets \_validity$
    \State $bidsPlaced\gets \_bidsPlaced$
    \State $biddingEnds\gets \_biddingEnd$
   \EndProcedure
  \end{algorithmic}

\end{algorithm}

\subsubsection{Full Track Scheme}
When a contractor places a tender (Algorithm \ref{alg:placeBidOS}), the first step is to place a smart contract elsewhere on the blockchain that contains only the tender data. Experiments on the Ropsten test blockchain showed that 5000 bits could be stored on the blockchain with a gas consumption that didn't get immediately rejected for being too expensive. This roughly translates to 700 words. It is suggested that the usually more verbose tender contracts are adapted for this format, perhaps by extracting the key information ($P_{TO}$) necessary for the tender. The data on this contract is available to anyone maintaining the blockchain, and so is protected through the encryption performed using the bidders bid-specific symmetric key sealed by $P_{TO}(SK_{Bidder})$ as discussed in step 4a in Section \ref{sec:OverallArchitecture}. 

Placing the tender for the auction involves using the auction smart-contract. The contractor passes in their ID, which should have been pre-agreed between the auction creator and the contractor, the address of their tender data on the blockchain, and then the components of the certificate that will have been given to them by the auction creator. This certificate should have been signed using the private key that corresponds to the request for tender's public key that was put into the contract upon creation. To offload some of the computational complexity of the contract, the parts of this certificate (the hashed message, and the v, r, and s values) are extracted on the client side. The request for tender will take the address of all placed tenders and store it in an array.

All tenders left on the contract are recorded, but the validity is determined using the certificate, the time that the tender was placed, and the record of bids placed. The time when the tender was placed is retrieved using the \textbf{\textit{now()}} function available on an Ethereum contract. The \textbf{\textit{now()}} function uses the time from epoch registered on the node executing the transaction. 

This timing can be considered trusted because of the timing consensus of nodes on the blockchain. Nodes are unable to mine blocks that have timestamps earlier than the parent blocks time stamp. Nodes are discouraged from placing blocks at an arbitrary amount of time ahead for the same principle – few nodes on the chain will be willing to put a block on the chain far ahead of their current time because they will be unable to place new blocks on top of it, thus stopping the block from getting picked up by a majority of nodes. Nodes are also discouraged from moving the time back because the challenge issued in the proof-of-work is orders of magnitude more difficult the shorter the interval between the current blocks timing and the parent blocks timing  \cite{wood2014ethereum}.

\begin{table*}[t]
\begin{minipage}[b]{0.30\textwidth}
\centering
\begin{tabular}{||c c c||} 
 \hline
 Contract & Time (secs) & Cost (gas) \\ [1ex] 
 \hline\hline
 1 & 351.35 & 352819 \\ 
 \hline
 2 & 122.41 & 352819  \\
 \hline
 3 & 156.5 & 352819  \\
 \hline
 4 & 112.55 & 352819 \\
 \hline
 5 & 108.48 & 352819 \\  
 \hline
 6 & 154.94 & 352819 \\ 
 \hline
 7 & 145.45 & 352819  \\
 \hline
 8 & 76.12 & 352819  \\
 \hline
 9 & 213.73 & 352819 \\
 \hline
 10 & 170.09 & 352819 \\ 
 \hline\hline
 Average & 147.761 & 892160 \\ 
 \hline
\end{tabular}
\caption{No State Scheme: Contract Deployment}
\label{table:NoStateSchContractDeployment}
\end{minipage}\hspace{2mm} 
\begin{minipage}[b]{0.33\textwidth}
\begin{tikzpicture}
\begin{axis}[
    ylabel={Gas Consumption in 10$^5$.},
    xmin=0, xmax=11,
    ymin=50000, ymax=200000,
    xtick={1,2,3,4,5,6,7,8,9,10},
    ytick={50000,75000,100000,125000,150000,175000,200000},
    legend pos=north west,
    ymajorgrids=true,
    grid style=dashed,
]
\addplot[
    color=blue,
    mark=square,
    ]
    coordinates {
    (1,156601)(2,156601)(3,156601)(4,156601)(5,156601)(6,156601)(7,156601)(8,156601)(9,156601)(10,156601)
    };
    \legend{Bid's Gas Consumption}
\end{axis}
\end{tikzpicture}

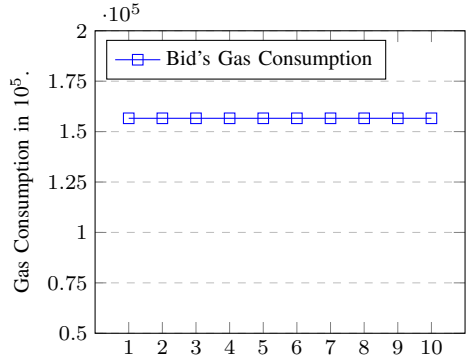
\captionof{figure}{Individual Bid's GAS usage (x 10$^5$) for Stateless Scheme}
\label{fig:StatelessContractGasUsage}
\end{minipage}\hfill 
\begin{minipage}[b]{0.33\textwidth}
\centering
\begin{tikzpicture}
  \begin{axis}
    [ylabel = {Time (Seconds)},
    boxplot/draw direction=y,
    xtick={1,2,3, 4, 5, 6, 7, 8, 9, 10},
    xticklabels={T1, T2, T3, T4, T5, T6, T7, T8, T9, T10},
    x tick label style={font=\footnotesize, text width=2.5cm, align=center}
    ]
    \addplot+[mark = *, mark options = {black}, 
    boxplot prepared={
      lower whisker=34,
      lower quartile=45,
      median=51,
      upper quartile=60.5,
      upper whisker=131,
      average=63.4
    }, color = red
    ]coordinates{};
    \addplot+[mark = *,mark options = {blue},
    boxplot prepared={
      lower whisker=21,
      lower quartile=25.5,
      median=42,
      upper quartile=47.75,
      upper whisker=65,
      average=40.1
    }, color = blue
    ] coordinates{};
    \addplot+[mark = *,mark options = {cyan}, 
    boxplot prepared={
      lower whisker=26,
      lower quartile=37.75,
      median=43.5,
      upper quartile=55.5,
      upper whisker=97,
      average=49.8
    }, color = Aquamarine
    ] coordinates{};
    \addplot+[mark = *, mark options = {Aquamarine}, 
    boxplot prepared={
      lower whisker=6,
      lower quartile=15.75,
      median=24.5,
      upper quartile=28.75,
      upper whisker=36,
      average=22.2
    }, color = Orange
    ]coordinates{};
    \addplot+[mark = *,mark options = {Orange}, 
    boxplot prepared={
      lower whisker=11,
      lower quartile=27,
      median=35.5,
      upper quartile=42.75,
      upper whisker=54,
      average=34.3
    }, color = OrangeRed
    ] coordinates{};
    \addplot+[mark = *,mark options = {OrangeRed}, 
    boxplot prepared={
      lower whisker=15,
      lower quartile=32.25,
      median=33.5,
      upper quartile=41.5,
      upper whisker=60,
      average=35.6,
      every box/.style={thin,solid,draw=Periwinkle,fill=white},
      every whisker/.style={Periwinkle, thin, solid},
      every median/.style={solid,Periwinkle,thin}
    }, color = Periwinkle
    ] coordinates{};    
    \addplot+[mark = *, mark options = {red}, 
    boxplot prepared={
      lower whisker=26,
      lower quartile=34.75,
      median=41.5,
      upper quartile=55,
      upper whisker=62,
      average=43.5,
      every box/.style={thin,solid,draw=Purple,fill=white},
      every whisker/.style={Purple, thin, solid},
      every median/.style={solid,Purple,thin}
    }, color = Purple
    ]coordinates{};
    \addplot+[mark = *,mark options = {blue}, 
    boxplot prepared={
      lower whisker=19,
      lower quartile=30.75,
      median=33,
      upper quartile=42.5,
      upper whisker=52,
      average=35.3,
      every box/.style={thin,solid,draw=CarnationPink,fill=white},
      every whisker/.style={CarnationPink, thin, solid},
      every median/.style={solid,CarnationPink,thin}
    }, color = CarnationPink
    ] coordinates{};
    \addplot+[mark = *,mark options = {green}, 
    boxplot prepared={
      lower whisker=8,
      lower quartile=26.5,
      median=39.5,
      upper quartile=48.25,
      upper whisker=183,
      average=50.1,
      every box/.style={thin,solid,draw=Maroon,fill=white},
      every whisker/.style={Maroon, thin, solid},
      every median/.style={solid,Maroon,thin}
    }, color = Maroon
    ] coordinates{};
       \addplot+[mark = *,mark options = {green}, 
    boxplot prepared={
      lower whisker=5,
      lower quartile=16.25,
      median=30.5,
      upper quartile=37.5,
      upper whisker=63,
      average=29.5,
      every box/.style={thin,solid,draw=JungleGreen,fill=white},
      every whisker/.style={JungleGreen, thin, solid},
      every median/.style={solid,JungleGreen,thin}
    }, color = JungleGreen
    ] coordinates{};
    \end{axis}
\end{tikzpicture}
\captionof{figure}{Bid timing for Stateless Scheme Over Ten Trial Tenders}
\label{fig:statelessSchemeTimiing}
\end{minipage}
\end{table*}


A bid is considered valid if the following hold ; 1) The contract is placed within time, 2) the certificate is verified to match the public key held by request for tender contract, and 3) the contractor is allowed to place more bids. If it fails any of these checks, it is still placed but flagged as an invalid bid. Note that the number of bids placed by the contractor is only incremented if it is being placed using a valid certificate, protecting the contractor from being locked out of the auction by a malicious party placing arbitrary numbers of bids using their identifier. 

\begin{algorithm}
\caption{Placing A Bid (Protected State Scheme)}\label{alg:placeBidPS}\Comment{Running on the local machine} 
\addtolength\linewidth{-4ex}
\small
\algnewcommand{\And}{\textbf{and}\xspace}
\begin{algorithmic}[1]
\Procedure{PlaceBid}{$id,data,msgHashed,v,r,s,$}
\State $validHash\gets verify(msgHash,v,r,s)$ 
\If{$ValidHash$}
\State \emph{bidValidity}$\gets$ \emph{ValidBid}(\emph{id})
\If{$bidValidity$}
\State $bidCount[id] += 1$
\EndIf
\State \emph{bid}$\gets$ new \emph{Bid}(\emph{id},\emph{data},\emph{bidValidity},\emph{bidsPlaced},\emph{biddingEnd})
\State $bidsPlaced.add(bid)$
\State \textbf{return} $bid$
\EndIf
\EndProcedure
\Procedure{ValidBid}{$id$}
\State $validTime\gets timeNow() < biddingEnd$
\State $allowedBid\gets bidCount[id] < limit$
\State \textbf{return} $validTime$ \And $allowedBid$
\EndProcedure
\Procedure{Bid}{\emph{\_id},\emph{\_data},\emph{\_validity},\emph{\_bidsPlaced},\emph{\_biddingEnd
}}
\State $id\gets \_id$ 
\State $data\gets \_data$
\State $validity\gets \_validity$
\State $bidsPlaced\gets \_bidsPlaced$
\State $biddingEnds\gets \_biddingEnd$
\EndProcedure
\end{algorithmic}
\end{algorithm}

The tender reference smart contract is then created and passed the id of the contractor, the address of the tender data, the validity of the contract, a copy of the array holding all of the addresses of the tenders placed before this one, and the auction end time. The address table is required to ensure the integrity of the array revealed by the auction creator – no bids could be intentionally erased from the array because the record of its existence will exist in the tender reference contracts.

Once the auction time has elapsed, the addresses can be requested from both the request for tender contract and the tender reference contracts. 

Algorithm \ref{alg:placeBidOS} performance indicates (Figure \ref{fig:FTContractGasUsage}) that altering the contract's state to hold the addresses increases the gas usage, and thus the cost, on each subsequent transaction. This is not desirable, particularly as failing contracts are also registered in the smart contract, which leaves the request for tender at risk of a denial of service attack.


\subsubsection{Protected Scheme}
Algorithm \ref{alg:placeBidPS} proposed an alternative schema to mitigate this potential denial of service attack. In this schema, the Request For Tender contract does not record bids if they fail the certificate check. We originally wanted to record every bid that was placed so the whole process would be recorded, and every attempted transaction would be visible. However, allowing anyone to place bids when each bid dramatically increases the amount of gas (and cost) necessary to place a new bid could leave our scheme open to a Denial of Service (DoS) attack. Although this scheme requires more gas initially, the first tender placed uses an amount of gas slightly higher than the third tender in the full tracking schema, given that it protects against DOS attacks from unauthorised parties, the gas hit is comparatively negligible.

Of course, in both of these proposals, the changing state of the Request For Tender contract does still lead to a dramatic increase in gas usage per transaction, regardless of whether or not those bids are intentionally malicious.

\begin{algorithm}\Comment{Running on the local machine} 
\caption{Placing A Bid (Stateless Scheme)}\label{alg:placeBidNS}
\addtolength\linewidth{-4ex}
\small
\algnewcommand{\And}{\textbf{and}\xspace}
\begin{algorithmic}[1]
\Procedure{PlaceBid}{$id,data,msgHashed,v,r,s,$}
\State \emph{bidValidity}$\gets$ \emph{ValidBid}(\emph{id},\emph{msgHashed},\emph{v},\emph{r},\emph{s})
\If{$bidValidity$}
\State $bidCount[id] += 1$
\EndIf
\State \emph{bid}$\gets$ new \emph{Bid}(\emph{id},\emph{data},\emph{bidValidity},\emph{bidsPlaced},\emph{biddingEnd})
\State \textbf{return} $bid$
\EndProcedure
\Procedure{ValidBid}{$id$}
\State $validHash\gets verify(msgHash,v,r,s)$ 
\State $validTime\gets timeNow() < biddingEnd$
\State $allowedBid\gets bidCount[id] < limit$
\State \textbf{return} $validHash$ \And $validTime$ \And $allowedBid$
\EndProcedure
\Procedure{Bid}{\emph{\_id},\emph{\_data},\emph{\_validity}}
\State $id\gets \_id$ 
\State $data\gets \_data$
\State $validity\gets \_validity$
\EndProcedure
\end{algorithmic}
\end{algorithm}

\subsubsection{Stateless Scheme} Algorithm \ref{alg:placeBidNS} provides an alternative by not storing the Tender Reference contracts in an array on the Request For Tender contract. This solves the increasing gas issue because we would no longer be changing the state of the Request For Tender contract. 
This would mean that a Tender Reference Contract would no longer need to hold information about previously placed bids and would also not require knowing when the tendering period ends because there would be no information to be queried from the Tender Reference Contract after the request for tender period lapses.

Instead, when a bid is placed, and the address is returned to the contractor, this address could then be given to the auction creator external to the transaction. The requirement would then be for the auctioneer to use a third part chain explorer to verify that the contract was made as the result of a transaction registered to the request for tender contract. The issue here would be that the third parties would have to be considered trusted and that they may, for non-malicious reasons, not have the required information (either from memory constraints necessitating that not all transactions are registered, or from missing the transaction altogether). However, if the contractor requires a certificate of acknowledgement from the auctioneer after they have given their tender-reference address, the auctioneer will not be able to refute that they received the tender.

\begin{algorithm}
\caption{Evaluating All Bids}\label{alg:evalBids}
\addtolength\linewidth{-4ex}
\small
\begin{algorithmic}[1]
\Procedure{MakeRequest}{$\_length,\_pubk,\_limit$}\Comment{Running on the local machine} 
\State $listOfBids \gets ReqBids()$
\For{\emph{bids} in \emph{listOfBids}}
\State \emph{validBid} $\gets${bids.getValidity()}
\If{validBid}
\State \emph{listOfValidBidDataAddresses}.add(validBid.getDataAddress()) 
\EndIf
\EndFor
\EndProcedure
\Procedure{ReqBids}{$\_length,\_pubk,\_limit$}\Comment{Running in the blockchain}
\State $afterAuction \gets timeNow() > biddingEnd$
\If{afterAuction}
\State \textbf{return} $bidsPlaced$
\EndIf
\EndProcedure
\end{algorithmic}
\end{algorithm}


\subsubsection{Bid Evaluation}
Algorithm \ref{alg:evalBids} is a retrieval algorithm that is only applicable to the schemas where the addresses are held in the state of the Request For Tender contract. The addresses would be retrieved from the contract after the request for the tender period has elapsed. The client application is used to interact with the blockchain will request the contract on the blockchain. Retrieving the information does not require any transactions and thus incurs no transaction cost.

The client application will receive a list of bids (in reality, this will be the addresses of the bids placed). These bids can then be queried for their validity and, once ascertained, the address of the actual tender data can be requested. The whole algorithm has a running time of O(n).

\section{Security and Operational Analysis}
\label{sec:SecurityAnalysis}
We now evaluate how well our proposal met the security and operation requirements as stipulated in Section III for each of the three variants of the open and transparent tendering framework, as discussed in Section IV. The table shows the extent to which each of the variants satisfies the stated requirements. 

\begin{table}[h]
 \centering
 \caption{Security and Operational Requirements Analysis for Three Variants.}
 \label{tab:SecOpeAnalysis}
 \resizebox{0.968\columnwidth}{!}{%
  \begin{tabular}{@{}cccc@{}}
   \toprule
   & \textbf{Full Track Scheme} & \textbf{Protected Scheme} & \textbf{Stateless Scheme} \\ \midrule
   \multicolumn{4}{c}{\cellcolor[HTML]{EFEFEF}\textbf{Blockchain Architecture}} \\
   \multicolumn{1}{l}{\textbf{Centralised}} & $\sq\sq\sq\sq\sq$ & $\sq\sq\sq\sq\sq$ & $\sq\sq\sq\sq\sq$\\
   \multicolumn{1}{l}{\textbf{Open}} & $\bl\bl\bl\bl\bl$ &$\bl\bl\bl\bl\bl$ & $\bl\bl\bl\bl\bl$\\
            \multicolumn{1}{l}{\textbf{Distributed}} & $\bl\bl\bl\bl\bl$ & $\bl\bl\bl\bl\bl$ & $\bl\bl\bl\bl\bl$\\
   \multicolumn{4}{c}{\cellcolor[HTML]{EFEFEF}\textbf{Security and Operational}} \\
   \multicolumn{1}{l}{\textbf{$R_1$}} & $\bl\bl\bl\bl\bl$ & $\bl\bl\bl\bl\bl$  & $\bl\bl\bl\bl\bl$ \\
   \multicolumn{1}{l}{\textbf{$R_2$}} & $\bl\bl\bl\bl\sq$ & $\bl\bl\bl\bl\sq$  & $\bl\bl\bl\bl\sq$\\
            \multicolumn{1}{l}{\textbf{$R_3$}} & $\bl\bl\bl\bl\bl$ & $\bl\bl\bl\bl\bl$ & $\bl\bl\bl\bl\bl$ \\
            \multicolumn{1}{l}{\textbf{$R_4$}} & $\bl\bl\bl\bl\sq$ & $\bl\bl\bl\bl\sq$  & $\bl\bl\bl\bl\bl$ \\
   \multicolumn{1}{l}{\textbf{$R_5$}} & $\bl\bl\bl\bl\sq$ & $\bl\bl\bl\bl\sq$  & $\bl\bl\bl\bl\bl$\\
            \multicolumn{1}{l}{\textbf{$R_6$}} & $\bl\bl\bl\bl\bl$ & $\bl\bl\bl\bl\bl$ & $\bl\bl\bl\bl\bl$ \\
 \bottomrule
  \end{tabular}
 }
\end{table}

As all of the proposed variants of the open and transparent tendering framework rely on open blockchain infrastructure, they all satisfy the open and distributed environment requirement. This requirement is imposed on the blockchain technology so open and fair evaluation of the activities on the blockchain can be conducted. 

All three of the schemes do not fully meet the $R_2$ requirement for the reason that if a bidding organisation shares the second half of the $P_{TO}(SK_{Bidder})$ (the first half is on the blockchain with the bid and second half remains with the bidding organisation) to the tendering organisations before the deadline, then, yes, the tendering organisation can read the bids. In all of the proposed schemes, we do not enforce that the key can only be shared after the deadline. This was to accommodate a bidding organisation's business processes and priorities. 

For $R_4$, the full track and protected schemes do not fully support the requirement. This is due to the exponential increase in the Gas costs when placing bids on the blockchain (Figures \ref{fig:PSContractGasUsage} and \ref{fig:StatelessContractGasUsage}), which could escalate beyond what can be considered a reasonable cost for business practices \ref{garcia2017optimized}. In theory, a malicious entity can place so many bids on a tender that it becomes economically too costly for genuine bidders to place a bid. However, from a practical point of view, the high number of bids necessary to make Gas cost extranomical would cost the malicious entity a huge sum of money. Furthermore, such a high number of bids would also be easily detected by the tendering organisation and any third party monitoring the tendering process. Therefore, such an attack might not be an attractive prospect for a maligned actor. 

\section{Conclusion and Future Research Directions}
\label{sec:Conclusions}
Openness and transparency are frequently discussed in the public service domain. Traditionally it has been difficult to build a transparent governance model. This was because it required an investment of both time and money from all stakeholders, especially the citizens. With the increasing adoption of e-government and open government initiatives, public opinion is in favour of developing innovative solutions that can increase openness and transparency in government activities with minimum cost to citizens. For citizens to be involved in monitoring the governance activities, they need efficient tools and intuitive assessment that gives clear results. To build such an environment, blockchain and smart contracts show great potential. In this paper, the government tendering process is implemented in the blockchain environment to provide an open and fair tendering scheme. Based on the proposed architecture, we put forward three variants that were then implemented on Etherium to show their applicability, Gas cost and computational performance. The main objective of the paper was to show that the tendering scheme can be made fully open, autonomous, fair and transparent using smart contracts. To this end, it was successful. 
There are two future research directions, 1) build additional government services on blockchains to increase openness and transparency and 2) enhancing the smart contract platform to be more feature rich, autonomous and supporting secure distributed execution. 

\addtolength{\textheight}{-12cm}   







\bibliographystyle{IEEEtran}
\bibliography{main}

\end{document}